\documentclass[twocolumn, amssymb, amsmath,nobibnotes,nofootinbib, aps, prl,showpacs]{revtex4}
\usepackage[dvips]{graphicx}
\begin{document}
\title{Metastability and magnetic memory effect in Ni-Mn-Sn  alloy}
\author{S. Chatterjee}
\author{S. Giri}
\author{S. Majumdar}
\email{sspsm2@iacs.res.in} 
\affiliation{Department of Solid State Physics, Indian Association for the Cultivation of Science, 2A \& B Raja S. C. Mullick Road, Jadavpur, Kolkata 700 032, India }
\author{S. K. De}
\affiliation{Department of Materials Sience,Indian Association for the Cultivation of Science, 2A \& B Raja S. C. Mullick Road, Jadavpur, Kolkata 700 032, India}
\begin{abstract}

Magneto-structural instability in the  ferromagnetic shape memory alloy of composition Ni$_2$Mn$_{1.4}$Sn$_{0.6}$ is investigated by transport and magnetic measurements. Large negative magnetoresistance is observed around the martensitic transition temperature (90-210 K). Both magnetization and magnetoresistance  data indicate that upon the application of an external magnetic field at a constant temperature, the sample attains a field-induced arrested state which persists even when the field is withdrawn. We observe an intriguing behavior of the arrested state that it can remember the last highest field it has experienced.  The field-induced structural transition plays the key role for the observed anomaly and the observed irreversibility can be accounted by the Landau-type free energy model for the first order phase transition.

\end{abstract}
\maketitle

The functional behavior of the shape memory alloys as actuators, magneto-mechanical transducers, switching devices, is related to the  structural instability known as  martensitic transition (MT). It is  defined as a  displacive, diffusionless first-order solid to solid phase transition from the high temperature austenite to the  low temperature martensite~\cite{mt}. Ferromagnetic shape memory alloys (FSMAs) combine shape memory effect and the bulk ferromagnetic behavior. MT, being key to the shape memory and some other metallurgic phenomena, has been under extensive research for over a century. The first order MT in Heusler based alloys, influenced by the disorder, often developed with a region of metastability with  austenite and the martensite coexisting together~\cite{kartha,sbroy, chatterjee}. MT is found to be highly sensitive to the external parameters like stress and magnetic field, and  magnetic field-induced strain\cite{shape, kainuma}, super-elasticity, magneto-caloric effect~\cite{imce, krenke}, giant magnetoresistance \cite{mr1, vishnu}  have been reported for many FSMAs. Materials with general formula Ni$_2$Mn$_{1+z}$X$_{1-z}$ (where X is an $sp$ element like, Sn, Sb, In etc.) are found to be quite useful for  possible multi-functional applications. 

\par
It is important  to understand the effect of magnetic field on the electronic and magnetic behavior of those FSMAs. This is  not only essential for the technological applications but can  provide fundamental insight of the phase separated scenario. In order to address the issue, we have performed detailed study of the magneto-transport and magnetic properties of the alloy Ni$_2$Mn$_{1.4}$Sn$_{0.6}$, which has been reported to show ferromagnetic shape memory effect and inverse magneto-caloric effect. The sample orders ferromagnetically below 330 K and  on cooling it undergoes MT from the cubic  phase to orthorhombic  phase around 180 K ~\cite{sutou} with a wide region of phase separation which extends from 90-210 K.  Our Study reveals large magnetoresistance (MR) around MT and unique magnetic behavior which go beyond the previous observations in case of ferromagnetic shape memory alloys. The resistivity and the magnetization show striking memory effect with respect to the applied magnetic field ($H$).

\par
The present investigation was carried out on polycrystalline sample of Ni$_2$Mn$_{1.4}$Sn$_{0.6}$ prepared by argon arc melting. The sample was characterized by  X-ray powder diffraction (XRD) and energy dispersive X-ray spectroscopy. The resistivity ($\rho$) of the sample was measured by usual four point method. MR (=$\frac{\rho(H)-\rho(0)}{\rho(0)}$) measurement was carried out using a superconducting magnet system in the transverse geometry ($H \perp$ current). Magnetization ($M$) was  measured   using a commercial Vibrating Sample Magnetometer (Cryogenic Ltd., UK).
 
\begin{figure}[t]
\centering
\includegraphics[width = 8 cm]{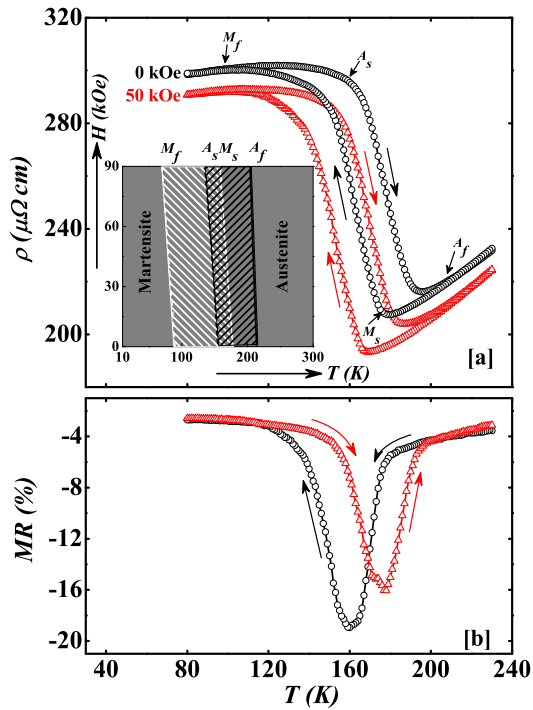}
\caption {(a) Resistivity as a function of temperature for both heating and cooling cycles (indicated by arrows) at zero and 50 kOe of applied fields for Ni$_2$Mn$_{1.4}$Sn$_{0.6}$ . The inset shows the $H-T$ phase diagram of the sample, with shaded area being the region of phase coexistence. (b) Magnetoresistance (MR) as a function of temperature for heating and cooling cycles for 50 kOe of field.}
\label{fig1}
\end{figure}

\par
Fig.~\ref{fig1} (a) shows $\rho$ versus temperature ($T$) behavior of the sample measured at $H$ = 0 and $H$ = 50 kOe  for both heating and cooling cycles. The clear thermal hysteresis around the first order MT is present for both zero field and 50 kOe runs. However, the MT is shifted to lower $T$ in presence of $H$.  In the  cooling run,  the  martensite  develops between temperatures $M_s$ and $M_f$, while austenite develops between $A_s$ and $A_f$ during heating.
In the $H-T$ phase diagram of the sample (inset of Fig.~\ref{fig1}(a)), the white and black shadings respectively denote the region of phase coexistence for cooling and heating legs. The average estimated shift of $M_s$ for 50 kOe of magnetic field is about 12 K. This produces large negative MR in the sample. Fig. ~\ref{fig1}(b) shows the plot of MR at 50 kOe versus $T$ for both heating and cooling legs. MR of about -16\% is observed in the sample for the heating leg, while on the cooling leg, MR is found to be around -19\%. It is clear that the large MR is only observed around the region of phase separation, which is approximately  the region of thermal hysteresis. The apparent difference of the magnitude of MR in the heating and cooling is due to the difference in the austenite and martensite fractions in the sample.

\par
\begin{figure}[t]
\centering
\includegraphics[width = 8 cm]{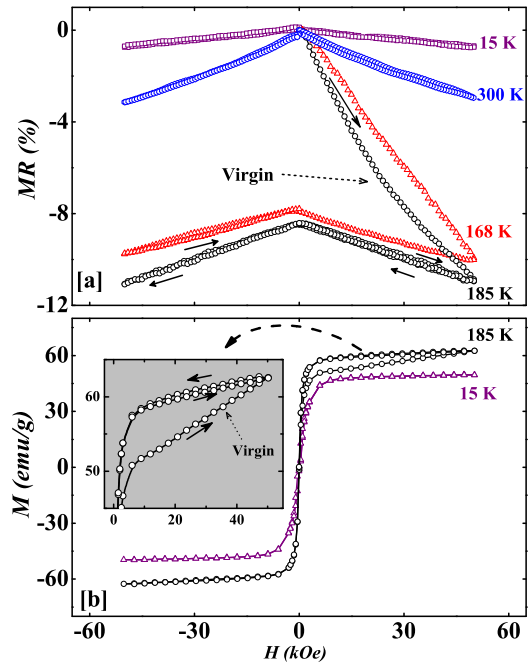}
\caption {(a) Magnetoresistance and (b) magnetization  as a function of field at different  constant temperatures. The sample was first zero field cooled to 10 K and then heated back to the respective constant temperatures for   MR and $M$ measurements. The data were collected for change of the magnetic field in the positive and negative quadrants. The inset in (b) represents the enlarged view of the magnetization data at 185 K.}
\label{fig2}
\end{figure}

In order to get a better look of the MR behavior, we  measured  $\rho$ as a function of field at different constant $T$ (Fig.\ref{fig2}(a)). The magnitude of MR is very small at 15 K, where the sample is in the pure martensite. However, larger negative MR is observed in the region of MT, which is in line with the $\rho(T)$  measurements performed at different fields. The interesting observation from the MR versus $H$ data is that in the region of metastability, the MR is highly irreversible with respect to the applied magnetic field. The large change in $\rho$ is observed when the field is applied first, however sample almost retains its low $\rho$  when the field is removed, and it persists for subsequent field cycling. This has been clearly depicted in Fig.~\ref{fig2}(a) where five loop field cycling is performed. For example, in case of  185 K isotherm, the initial application of 50 kOe of field (virgin leg) produces about -11\% of MR, while in the subsequent field removal and  application (even in the negative $H$ quadrant) do not produce much effect on $\rho$, only a mere 2.5\% change is observed. This type of field arrested state is only restricted across the MT and no indication of arrested state is observed above and below the MT (15 K and 300 K respectively in Fig.~\ref{fig2}(a)). Notably, the MR does not show any signature of saturation till 90 kOe, and the arrested state can be created even at 90 kOe (not shown here).

In a ferromagnetic alloy, it is expected that the MR anomaly has some magnetic origin. To probe it, we  measured isothermal $M$ as a function of field at different constant temperatures. The initial magnetization leg (virgin leg: 0 to 50 kOe) at 185 K shows larger slope beyond the point of technical saturation as compared to the return leg. On subsequent field cycling, very similar to the MR behavior, $M$ traces the  return leg  with almost zero coercivity. The virgin loop remains well outside (below) the subsequent field-cycling loops (see inset of figure 2 (b)). It appears that by the application of 50 kOe of field, the sample has been arrested to a  state with a very soft ferromagnetic character. Similar field-induced arrested state is also reported for Ni-Mn-X based alloys~\cite{kainuma, vishnu, koyama}. For the present sample, we find that there is a threshold field of about 1 kOe for the production of this arrested state.

\par
It is tempting to know the character of this field-induced arrested state and what happens to the resistivity if the temperature is varied after this state is achieved. For this purpose, we performed $\rho$ measurement in the following protocol (see figure 3(a)). The sample was first cooled in zero magnetic field from 240 K to 80 K (point $c$ to $d$ in figure 3) and heated back to 180 K (point $e$). At this point, 50 kOe of field was applied, and $\rho$ drops down to point $f$. Immediately, field was removed and the sample was allowed to heat up. $\rho$ followed a completely different path ($gh$) and eventually joined up with the zero field heating curve at the point $h$. It clearly depicts that the application and subsequent removal of $H$ at a certain point inside the thermal hysteresis region produces a state which is different from the zero field state as far as the $T$ dependence of resistivity is concerned. However, eventually it approaches the zero field virgin state. 

\begin{figure}[t]
\centering
\includegraphics[width = 8 cm]{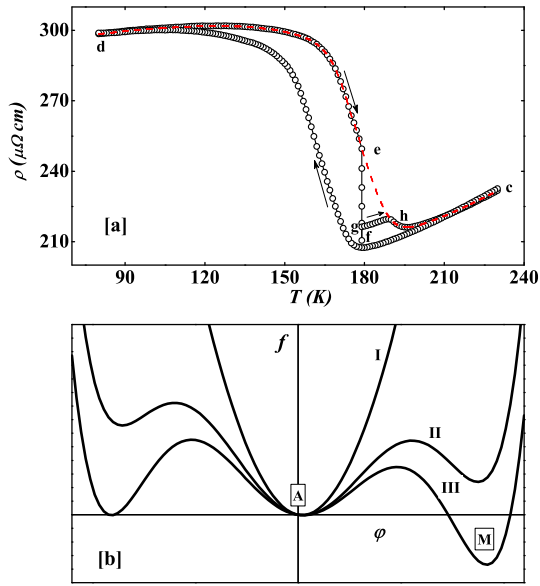}
\caption {(a) Resistivity as a function of temperature with a protocol discussed in the text. (b) Landau type free energy diagram (see text) for the first order phase transition for decreasing temperature (I to III) in presence of field. ``A'' and ``M'' denote austenite and martensite minima respectively. It is to be noted that the free energy profile is asymmetric with respect to the $\varphi$ = 0 line due in presence of field.}
\label{fig3}
\end{figure}

\begin{figure}[t]
\centering
\includegraphics[width = 8 cm]{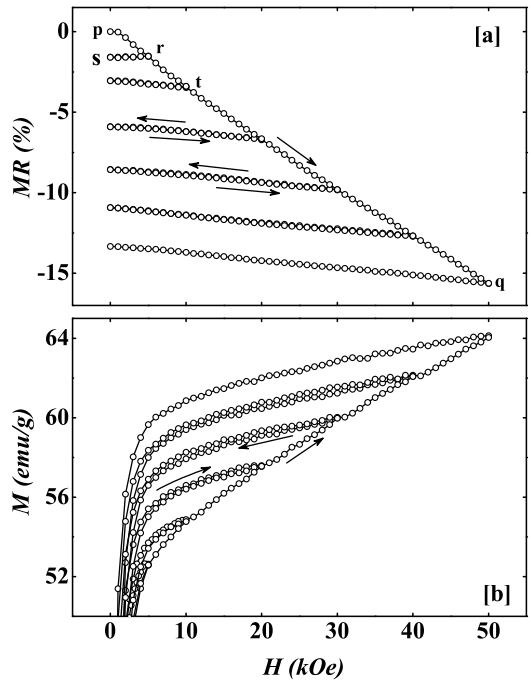}
\caption {(a) Magnetoresistance and (b) Magnetization as a function of field recorded at 180 K. The measurement was performed by returning the field back to zero from different maximum fields ($H_m^i$).}
\label{fig4}
\end{figure}

The noteworthy observation is the magnetic memory effect as depicted in fig~\ref{fig4}. We have seen that the sample goes to a different electronic and magnetic state on application of a field and the state is retained on removal of the field. The question is whether the nature of the arrested state depends upon $H$ or not.  To shed light on it, the sample was first zero field cooled to 10 K and then heated back to 180 K. At 180 K, $H$ was increased from zero to a certain value $H_m$ and then removed gradually. This was repeated for increasing values of $H_m$ (say, $H_m^1$, $H_m^2$...$H_m^i$...)  staring from  5 kOe to 50 kOe. As expected, once a certain field $H_m^i$ was applied, the sample retained the lower resistivity value even after the field was brought back to zero. In the next cycle, when the field was ramped up to the next higher value of $H_m^{i+1}$, $\rho$ followed the same ramping down path till the previous applied field $H_m^i$ was reached. Beyond  $H_m^i$, $\rho$ followed the master path, $pq$. The master path can be obtained if $H$ is ramped up monotonously to its highest value.  The minor paths (like, $sr$) intercepted the master path on their respective maximum exposed field values. This clearly shows that the field-induced state remembers the value of the last highest field ($H_m^i$) it has experienced. This is an  example of magnetic memory found in this alloy, where different electronic and magnetic states can be created by varying the value of $H_m$.  It has been depicted in figure 4: for example, the field was first ramped up to 5 kOe  (point $p$ to $r$) and then ramped down to zero ($r$ to $s$). When the field was again ramped up from zero to the next field value of 10 kOe, $\rho$ followed the path $s$ to $r$ till the previous highest experienced field (here 5 kOe) was reached (point $r$) and after that it followed the master path ($pq$) to reach the point $t$ corresponding to $H_m$ = 10 kOe. Similar thing happens for higher and higher values of $H_m$ (even observed for $H_m$ = 90 kOe). The  memory effect is also seen in the $M$ versus $H$ measurement at 180 K as depicted in Fig.~\ref{fig4}(b). 

\par
Magnetic field-induced arrested state and magnetic memory effect  has been observed in systems like  frozen ferro-fluids, interacting nanoparticles, metal-insulator multilayer. However, in all those cases the physics behind it is the spin-glass like dynamics~\cite{sun}.  We do not see any signature of glassiness in the sample, for example, the ac susceptibility does not show any change in the martensitic transition temperature with frequency. The more probable scenario here is the magnetic field-induced first order phase transition~\cite{sbr}. The ability of the present sample to remember the field from where it has been backtracked is referred as {\it return point memory effect}~\cite{sethna, ortin} and it can take place  as a function of stress or temperature in addition to $H$. This is connected to the metastability of the system in the vicinity of a disordered induced first order transition~\cite{sethna}.  Recently, X-ray diffraction measurement has confirmed field-induced first order transition in  Ni-Mn-Sn alloy, where  martensite fraction transformed into the austenite by the application of $H$~\cite{koyama}. When a magnetic field is applied, some martensite fraction  transforms into asutenite, which has a lower resistivity  than martensite resulting large negative MR. The martensite has lower magnetic moment than austenite, as a result the effect of the transformation is also visible in $M$. When the field is removed the system does not recover the martensite fraction. This irreversible nature of the field-induced transition was found in  XRD measurement by Koyama {\it et al.}~\cite{koyama}.

\par
Phenomenologically, Landau type of free energy expression can been used to describe the first order MT~\cite{falk, huo}, where free energy and the order parameter($\varphi$) are related as $f(\varphi) = a_2\varphi^2 -a_4\varphi^4 + a_6\varphi^6 -\gamma\varphi$  with $a_i$'s  are the  parameters for the particular material, and $\gamma$ is a term proportional to the externally applied field. For $T$-dependent transition, $a_2 \sim a_2^0(T-T_0)$, and the free energy curves for different $T$ are plotted in  Fig.~\ref{fig3}(b). Across the region of transition, the martensite and the austenite  coexist in  different free energy minima separated by energy barrier. Thermal irreversibility in various physical quantities arise due to the fact that one of the states becomes metastable (supercooled or superheated state) and the system remains trapped there even when $T$ crosses the critical point.

\par
Similar situation can arise when the transition takes place as a function of $H$ rather than $T$. Since field favors the formation of austenite, application of $H$ is equivalent to the increase of temperature. $H$ changes the free energy pattern as well and allow redistribution of  the relative  fraction of product and parent phases. Suppose, the system is in a particular state at temperature $T_0$ and field $H_0$,  where both phases are coexisting with austenite and martensite fractions being respectively, x$_A$ and x$_M$ = 1- x$_A$. This state can be represented by curve (III) of free energy plot  shown in  Fig.~\ref{fig3}(b). As we  increase $H$  isothermally, the free energy curve will change to  curve (II). Here, the martensitic minimum lifts up to higher value and the barrier ($\sigma$ =$\sigma(T, H)$ ) between the parent and product phases decreases, allowing the formation of austenite. The new phase fractions becomes x$_A'$= x$_A$ + $\delta$x$_A$, x$_M'$ = (1- x$_A$ - $\delta$x$_A$). The fraction, $\delta$x$_A$, transformed from martensite to austenite,  depends upon the change in $H$. When the field is brought back to $H_0$, the free energy curve return back to its previous  field patter (from curve II to curve III). However, the free energy barrier does not allow the martensite to regain its transformed fraction, and we get a state with arrested (metastable) austenite fraction, $\delta$x$_A$. This state will bear the memory of the highest field, $H_m$, it has exposed to by virtue of the amount of $\delta$x$_A$ it contains. The arrested  state remains arrested until the barrier height is changed through the change in  $T$ or $H$. If $T$ is kept constant, and $H$ is increased from $H_0$, nothing will happen until the last highest exposed field is crossed. Further increase of $H$ (beyond $H_m$), however produces more amount of austenite  and the system attains a different x$_A$, x$_M$ ratio through further reduction of the free energy barrier. It is to be noted that different field induced arrested states can be created by any arbitrary applied field.  This is in contrast with the  arrested states in Gd$_5$Ge$_4$ and Mn-site doped manganites~\cite{hardy}, where field-induced transitions take place only at certain values of $H$.  

\par
In conclusion, the sample Ni$_2$Mn$_{1.4}$Sn$_{0.6}$ shows reasonably large magnetoresistance around the region of first order martesnitic transition. MR is highly irreversible with respect to the applied magnetic field. A nice magnetic memory effect is observe, as the sample remembers the last highest field that it was exposed to. This can be exploited for important device related applications in future.

\end{document}